\documentstyle[twocolumn,aps,epsfig]{revtex}

\begin{document}

\title{
Pseudorapidity and centrality dependence of the collective flow of
charged particles in Au+Au collisions at $\sqrt{s_{_{NN}}} =$ 130
GeV }
\author {
B.B.Back$^1$, M.D.Baker$^2$, D.S.Barton$^2$, R.R.Betts$^6$,
R.Bindel$^7$, A.Budzanowski$^3$, W.Busza$^4$, A.Carroll$^2$,
M.P.Decowski$^4$, E.Garcia$^6$, N.George$^1$, K.Gulbrandsen$^4$,
S.Gushue$^2$, C.Halliwell$^6$, J.Hamblen$^8$, C.Henderson$^4$,
D.Hofman$^6$, R.S.Hollis$^6$, R.Ho\l y\'{n}ski$^3$, B.Holzman$^2$,
A.Iordanova$^6$, E.Johnson$^8$, J.Kane$^4$, J.Katzy$^{4,6}$,
N.Khan$^8$, W.Kucewicz$^6$, P.Kulinich$^4$, C.M.Kuo$^5$,
W.T.Lin$^5$, S.Manly$^{8}$,  D.McLeod$^6$, J.Micha\l owski$^3$,
A.Mignerey$^7$, R.Nouicer$^6$, A.Olszewski$^{3}$, R.Pak$^2$,
I.C.Park$^8$, H.Pernegger$^4$, C.Reed$^4$, L.P.Remsberg$^2$,
M.Reuter$^6$, C.Roland$^4$, G.Roland$^4$, L.Rosenberg$^4$, J.
Sagerer$^6$, P.Sarin$^4$, P.Sawicki$^3$, W.Skulski$^8$,
S.G.Steadman$^4$, P.Steinberg$^2$, G.S.F.Stephans$^4$,
M.Stodulski$^3$, A.Sukhanov$^2$, J.-L.Tang$^5$, R.Teng$^8$,
A.Trzupek$^3$, C.Vale$^4$, G.J.van Nieuwenhuizen$^4$,
R.Verdier$^4$, B.Wadsworth$^4$, F.L.H.Wolfs$^8$, B.Wosiek$^3$,
K.Wo\'{z}niak$^{3}$, A.H.Wuosmaa$^1$, B.Wys\l ouch$^4$\\ $^1$
Physics Division, Argonne National Laboratory, Argonne, IL
60439-4843\\ $^2$ Chemistry and C-A Departments, Brookhaven
National Laboratory, Upton, NY 11973-5000\\ $^3$ Institute of
Nuclear Physics, Krak\'{o}w, Poland\\ $^4$ Laboratory for Nuclear
Science, Massachusetts Institute of Technology, Cambridge, MA
02139-4307\\ $^5$ Department of Physics, National Central
University, Chung-Li, Taiwan\\ $^6$ Department of Physics,
University of Illinois at Chicago, Chicago, IL 60607-7059\\ $^7$
Department of Chemistry, University of Maryland, College Park, MD
20742\\ $^8$ Department of Physics and Astronomy, University of
Rochester, Rochester, NY 14627\\ }

\date{\today}
\maketitle

\begin{abstract}\noindent
This paper describes the measurement of collective flow for
charged particles in Au+Au collisions at $\sqrt{s_{_{NN}}} =$ 130
GeV using the PHOBOS detector at the Relativistic Heavy Ion
Collider (RHIC). An azimuthal anisotropy is observed in the
charged particle hit distribution in the PHOBOS multiplicity
detector. This anisotropy is presented over a wide range of
pseudorapidity ($\eta$) for the first time at this energy. The
size of the anisotropy (v$_{2}$) is thought to probe the degree of
equilibration achieved in these collisions. The result here,
averaged over momenta and particle species, is observed to reach
7\% for peripheral collisions at mid-rapidity, falling off with
centrality and increasing $|\eta|$. Data are presented as a
function of centrality for $|\eta|<1.0$ and as a function of
$\eta$, averaged over centrality, in the angular region {\mbox
-5.0$<\eta<$5.3}. These results call into question the common
assumption of longitudinal boost invariance over a large region of
rapidity in RHIC collisions.

\end{abstract}



The study of collective flow in non-central ultra-relativistic
heavy ion collisions is important because it can provide
information on the initial spatial anisotropy of the reaction
zone, conditions present in early stages of the collision
\cite{r_r_review,sorge,kolb_soll_heinz,teaney_shuryak} and the
degree of thermalization attained during the evolution of the
collision \cite{kolb_sollfrank_heinz_2,voloshin_poskanzer_1}.
In addition, collective flow can affect other measurables of
interest, such as two particle correlation functions and the
slopes of transverse momentum distributions
\cite{r_r_review,heiselberg_levy}. Effects of collective flow have
been observed in nuclear collisions over a wide range of collision
energies and
species\cite{r_r_review,ollitrault,ags_results_1,ags_results_2,sps_results_1,sps_results_2,sps_results_3,star1,star2,phenix}.

Hydrodynamic models, which assume local thermal equilibrium at all
points, are generally thought to predict maximal flow.  Such
models are fairly successful at reproducing the mid-rapidity flow
results at RHIC for the more central events and lower transverse
momenta \cite{Heinz,hirano}.  This implies substantial early
equilibration in these collisions.  Hydrodynamic models predict
roughly constant flow over a broad pseudorapidity region  either
through a full three-dimensional calculation \cite{hirano} or by
assuming longitudinal boost invariance.  This paper provides an
additional constraint on such models.

The analysis presented here is based on data taken between June
and September, 2000, during the first RHIC physics run.  Results
are presented  for Au+Au collisions at $\sqrt{s_{_{NN}}} =$ 130
GeV as a function of centrality over a narrow range of
pseudorapidity ($\eta$), while minimum bias results are shown over
a large $\eta$ range (-5.0$<\eta<$5.3).



The PHOBOS detector employs silicon pad detectors to perform
tracking, vertex detection and multiplicity measurements. Details
of the setup and the layout of the silicon sensors can be found
elsewhere\cite{phobos1,phobos2,phobos3}. For this running period,
detector components relevant for this analysis included the first
six layers of both silicon spectrometer arms (SPECN and SPECP),
the silicon vertex detector (VTX), the silicon octagonal
multiplicity detector (OCT), three annular silicon ring
multiplicity detectors on each side of the collision point (RN,
RP), and two sets of scintillating paddle counters (PN, PP).

Collisions with a coincidence of two or more signals in each of
the PN and PP counters were selected by the trigger.  This sample
corresponded to $86\pm3$\% of the total inelastic Au+Au cross
section. The centrality determination for the triggered events was
based on a truncated mean of the deposited energy in the paddle
counters. This variable is proportional to the number of particles
hitting these counters and is monotonically related to the number
of participants, N$_{part}$. More details on the event triggering
and centrality determination can be found
elsewhere\cite{phobosprl}. Adjustments in the procedures used
previously were made to take into account the fact that data used
in this analysis came from an offset fiducial volume and included
periods with different magnetic field settings.


Monte Carlo (MC) simulations of the detector performance were
based on the Hijing event generator \cite{hijing} and the
GEANT~3.21 \cite{geant} simulation package, folding in the signal
response for scintillator counters and silicon sensors.


The anisotropy of the azimuthal distribution of charged particles
traversing the detector formed the basis for this flow analysis.
Uniform and symmetric acceptance was beneficial in terms of
sensitivity to the flow signal. This led to the requirement that
the primary collision vertex fall within an 8 cm fiducial region
centered at -34 cm from the nominal interaction point, in a
uniform and symmetric part of the OCT subdetector.

 The position of the primary collision vertex was determined on an
event-by-event basis by fitting for the optimal intersection point
of the straight tracks reconstructed in the first six planes of
each of the two spectrometer arms. In addition, the vertex
position along the beam, z$_{vtx}$, was required to be consistent
with the vertex position as determined by an independent algorithm
to avoid pathological vertex reconstructions and reduce the
potential for systematic effects. The second algorithm determined
z$_{vtx}$ as the z position of the maximum of the azimuthally
averaged hit density in the OCT subdetector. Finally, the
reconstructed transverse vertex position was required to be within
2$\sigma$ of the average position (beam orbit).

\begin{figure}[t]
\centerline{ \epsfig{file=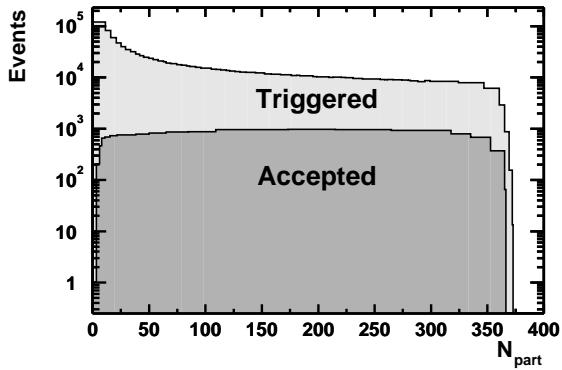,width=8cm} }
\caption{The event distribution as a function of N$_{part}$ for
triggered events (upper curve) and for data accepted for use in
the final analysis (lower curve).} \label{2}
\end{figure}
Trigger selection yielded $1.37\times 10^{6}$ events.  A total of
13,644 events survived the vertex cuts described above, yielding a
global acceptance for events used in the analysis of $\sim$1\%.
The number of triggered and accepted events are shown as a
function of N$_{part}$ in Fig.~\ref{2}. The vertex position
resolution for accepted events was estimated from simulated data
to be $\sim$1~mm in x, $\sim$1~mm in y and $\sim$3~mm in z.


The raw data for this analysis came in the form of energy
depositions in individual detector pads, known as hits.  The hit
energies were adjusted for variations in silicon thickness and
converted to dE/dx using the expected path length, assuming each
hit to come from a particle emanating from the reconstructed event
vertex.  Pads with energy depositions greater than 0.625 of the
peak of the minimum ionizing particle distribution (i.e., $>$50
keV) were taken to represent points of charged particle transit
and were used in the analysis. At the ends of the octagon, where a
single particle track often passes through more than one detector
pad, the energy signals in adjoining pads consistent with being
from a single track were added together.
To avoid biases introduced by malfunctioning pads, the signals
from such pads were replaced with signals from corresponding
mirror image pads in $\eta$, making use of the fact that the data
were taken during symmetric collisions. This was done for less
than 3\% of pads in the OCT and less than 1\% of the pads in RN
and RP. The position of each hit was smeared randomly with a flat
distribution within its hardware pad boundaries and mapped into
$\eta - \phi$ coordinates. The analysis was done on an
event-by-event basis in $\eta - \phi$ space.


The strength of the flow is given by the n$^{th}$ Fourier
coefficient of the particle azimuthal angle distribution,
\begin{equation}
 \frac{dN}{d(\phi-\psi_{2})}
\sim 1 + \sum_{n}2{\rm v}_{n}cos[n(\phi-\psi_{2})]
\end{equation}
where $\psi_{2}$ is the event plane angle, which is the best
estimate of the azimuthal reaction plane angle, $\psi_{R}$,
defined by the impact parameter and the beam axis. This analysis
was confined to n=2, the so-called elliptic flow. It generally
followed the scheme proposed by Poskanzer and Voloshin
\cite{pandv} and was based on the correlation of hits in one part
of the detector, known as a subevent, with the event plane angle
as determined by hits in a different part of the detector (a
different subevent). $\psi_{2}$ in a given subevent, `a', was
determined by
\begin{equation}
\psi_{2}^{a}=\frac{1}{2}tan^{-1}\left( \frac{\sum_{i}
w_{i}sin(2\phi_{i})}{\sum_{i} w_{i}cos(2\phi_{i})}\right),
\end{equation}
where the weights, $w_{i}$, were selected to maximize reaction
plane resolution by adjusting for acceptance and occupancy effects
as described below. The sums ran over all hits in subevent a.

The $\psi_{2}$ distribution as a function of azimuthal angle
should be flat in the absence of detector effects. Structures in
the raw $\phi$ hit and $\psi_{2}$ distributions were understood
qualitatively from studies of simulated data as coming from gaps
between sensors, channel-to-channel differences in $\phi$ phase
space coverage, and the asymmetric production of background
particles. These effects were removed (and the $\psi_{2}$
distribution flattened) in each $\eta$ annulus through the use of
individual hit weights, w$_{i}^{a}$, which were proportional to
the inverse of the average number of hits in each pad. Residual
effects due to the variation in the vertex position were absorbed
in the systematic error.  It should be noted that the final
results of the analysis were insensitive to this weighting and
that results with no acceptance weighting  were consistent with
the observations reported here.

\begin{figure}[h]
\centerline{\epsfig{file=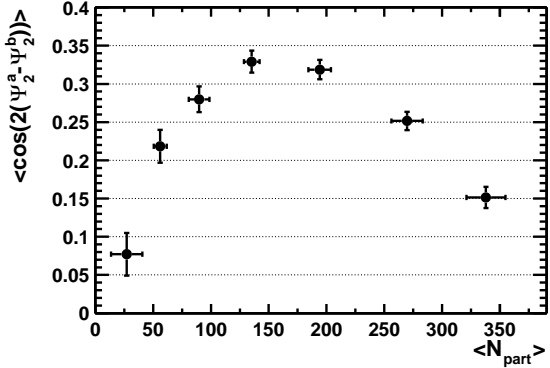,width=8cm}}
\caption{Reconstructed event plane angle correlation between
subevents OCT- and OCT+ as a function of centrality bin. This
quantity is directly related to the event plane resolution, as
shown in equation (5).} \label{6}
\end{figure}

The finite pad size in the detector led to an occupancy-dependent
hit saturation that reduced sensitivity to flow.  This effect was
accounted for in the analysis by weighting the hits in a given
$\eta - \phi$ bin by the average number of tracks per hit pad, or
occupancy, calculated individually in different sections of the
detector.
The occupancy was determined on an event-by-event basis from the
number of occupied (N$_{occ}$) and unoccupied (N$_{unocc}$) pads
in each section. The occupancy weight in a given section was
determined assuming a Poisson statistical distribution as
\cite{multdists}
\begin{equation}
Occ(\eta,\phi)=\frac{\mu}{1-e^{-\mu}},
\end{equation}
where $\mu$=ln(1 + N$_{occ}$/N$_{unocc}$) is the average number of
tracks per pad. This occupancy was used in concert with the
acceptance weight to produce the overall weight,
\begin{equation}
{\rm w}_{i}={\rm w}^{a}_{i}Occ(\eta_{i},\phi_{i}),
\end{equation}
which was used in the determination of $\psi_{2}$ in equation (2).
The event plane resolution, R, was calculated separately for each
centrality bin using the subevent technique\cite{pandv} as
\begin{equation}
R = \sqrt{<cos[2(\psi_{2}^{a}-\psi_{2}^{b})]>},
\end{equation}
where the superscripts denote separate subevents within a given
event and the averaging was done over all accepted events in a
given centrality bin. In this analysis, equal multiplicity
subevents were defined by dividing the event into two separate
$\eta$ regions, OCT-, which extended from -2.0 to -0.1 in $\eta$,
and OCT+, which encompassed $\eta$ between 0.1 and 2.0. The gap
between the two angular ranges in the OCT subdetector was
introduced to reduce effects due to short-range non-flow
correlations between hits in different subevents.  The final
v$_{2}$ determination was found to be robust against the choice of
subevent used in the evaluation of the event reaction plane so
long as the chosen subevent contained sufficient statistics that
the reaction plane resolution could be well determined.  In
addition, the gap width between subevents was varied from 0.2 to
1.0 in $\eta$ and the change in v$_{2}$ was incorporated in the
final systematic error for the analysis. The correlation between
the event planes in the OCT- and OCT+ subevents is shown in
Fig.~\ref{6}.

\begin{figure}[h]
\centerline{ \epsfig{file=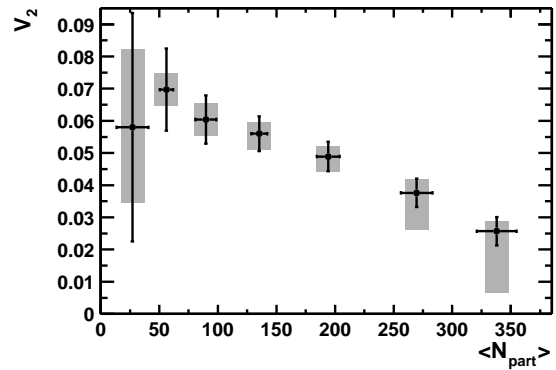,width=8cm} }
\caption{Fully corrected measurement of elliptic flow, v$_{2}$, as
a function of the number of participants for $|\eta|<$1.0.  The
black error bars represent the 1$\sigma$ statistical errors and
the gray bands give a measure of the systematic error for each
point at $\sim$90\% confidence level. } \label{7}
\end{figure}

The observed, resolution and occupancy corrected value of
 v$_{2}$, v$_{2}^{obs}$, was calculated in bins of
 centrality and $\eta$
 from the $\eta -\phi$ hit map using
\begin{equation}
{\rm v}_{2}^{obs}=\langle \frac{<w_{i}cos[2(\phi - \psi_{2})]>}{R}
\rangle,
\end{equation}
where the averaging in the numerator was done over the hits in one
event and the averaging of the fraction was done over all events
in the given centrality or $\eta$ bin using the appropriate value
of R for each event.

In this analysis, v$_{2}^{obs}$ was calculated in several regions
of the detector using the $\phi$ position of the hits in the $\eta
- \phi$ space. Two regions at high $\eta$ corresponded to areas
covered by the ring detectors, RN and RP, and encompassed
-5.0$<\eta<$-3.0 and 3.5$<\eta<$5.25, respectively. In the
mid-$\eta$ range, the regions covered by the OCT subdetector(OCT-
and OCT+) were used. For the determination of v$_{2}$ in the
positive (negative) $\eta$ region of the detector, OCT- (OCT+) was
used as the subevent region to evaluate $\psi_{2}$.  Fiducial cuts
in $\eta$ were used to avoid acceptance edge effects on the flow
signal.

\begin{figure}[t]
\centerline{ \epsfig{file=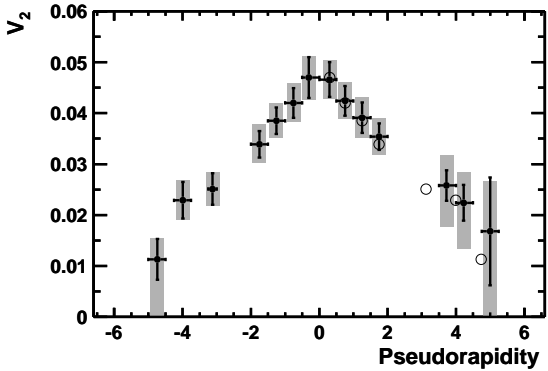,width=8cm} }
\caption{Elliptic flow,
averaged over centrality, as a function of $\eta$. The black error
bars represent 1$\sigma$ statistical errors and the gray bands
represent the systematic error at $\sim$90\% confidence level.
The points on the negative side are reflected about $\eta$=0 and
shown on the positive side as open circles. } \label{8}
\end{figure}

Even after resolution correction, Monte Carlo simulations showed a
residual suppression of the flow signal from background particles
carrying no flow information, the so-called non-flow background.
This effect was studied in detail using simulated data with a
known amount of flow.
By comparing the output resolution corrected flow signal to the
input flow signal for many samples of simulated data with
different forms and magnitudes of input flow (v$_{2}$),
suppression correction factors, C, were determined for each bin of
centrality and $\eta$ in the analysis.  The suppression factors
were found to be independent of the assumed flow magnitude and its
form. Furthermore, as a function of $\eta$, the correction is a
constant 12\% in the OCT and ranges from 15-30\% in most of the
rings, where the backgrounds are higher. As a function of
centrality the correction is a flat 12\%. The final corrected
value of v$_{2}$ was determined by
\begin{equation}
{\rm v}_{2}(\eta,centrality)=\frac{{\rm
v}_{2}^{obs}(\eta,centrality )}{C(\eta)}.
\end{equation}
This quantity, averaged over the region \mbox{-1.0$<\eta<$1.0}, is
presented in Fig.~\ref{7} as a function of the number of
participants, N$_{part}$.
The $\eta$ dependence of v$_{2}$, event averaged over centrality,
is shown in Fig.~\ref{8}.  For the sample of accepted events
entering this plot, $<{\rm N}_{part}>$=191.  This is substantially
more than the overall $<{\rm N}_{part}>$=98 for minimum bias
events in Hijing, reflecting the trigger inefficiency and the
centrality bias in our vertex reconstruction.

Numerous sources of systematic error were investigated, including
effects due to the energy cut, hit merging, subevent definition,
knowledge of the beam orbit relative to the detector, shape of the
dN/d$\eta$ distribution, vertexing algorithm, transverse vertex
cuts, magnetic field configuration and suppression correction
determination. The effect of these sources depended both on $\eta$
and centrality. In general, the systematic error arising from each
source was determined by conservatively varying that specific
aspect of the analysis (or several aspects in concert)  and
quantifying the change in the final v$_{2}$ result as a function
of $\eta$ and centrality. The individual contributions were added
in quadrature to derive the 90\% confidence level error shown as
the gray band in Figs.~\ref{7} and~\ref{8}. In addition to the
procedure described above, the systematic error on the lower side
of some points in Figs.~\ref{7} and~\ref{8} was increased to
reflect the reduced sensitivity of the analysis to the low level
of flow in those bins as determined from Monte Carlo studies.

These results represent the first measurement of v$_{2}$ as a
broad function of $\eta$ at RHIC and agree with the mid-rapidity
measurements of v$_{2}$ from the STAR \cite{star1} and PHENIX
\cite{phenix} collaborations. The extended angular coverage
portrayed in Fig.~\ref{8} clearly shows a systematic drop in the
magnitude of v$_{2}$ with $|\eta|$. This drop could be due, in
part, to correlations not associated with the reaction plane, a
drop in $\langle p_{T} \rangle$ due to dynamics or kinematics
\cite{kolb}, substantial directed flow, or a change in the
underlying mechanism of particle production as a function of
$\eta$. Since these data appeared in a preliminary form
\cite{IP_QM01}, theoretical efforts have had only limited success
in reproducing the $\eta$ dependence \cite{hirano,QGSM}. These
results call into question the common assumption of longitudinal
boost invariance over a large region of rapidity in RHIC
collisions. Further work is needed to develop a clear,
three-dimensional picture of these collisions.

Acknowledgements: We acknowledge the generous support of the C-A
Division and Chemistry Departments at BNL. We thank Fermilab and
CERN for help in silicon detector assembly. We thank the MIT
School of Science and LNS for financial support. This work was
partially supported by US DoE grants DE-AC02-98CH10886,
DE-FG02-93ER40802, DE-FC02-94ER40818, DE-FG02-94ER40865,
DE-FG02-99ER41099, W-31-109-ENG-38. NSF grants 9603486, 9722606
and 0072204. The Polish groups were partially supported by KBN
grant 2 P03B 04916. The NCU group was partially supported by NSC
of Taiwan under contract NSC 89-2112-M-008-024.



\end{document}